# An ultrafast reconfigurable nanophotonic switch using wavefront shaping of light in a nonlinear nanomaterial

Tom Strudley[1], Roman Bruck[1], Ben Mills[2], Otto L. Muskens[1]*

[1] Physics and Astronomy, Faculty of Physical Sciences and Engineering, University of Southampton, Highfield, Southampton SO17 1BJ, United Kingdom.

[2] Optoelectronics Research Centre, Faculty of Physical Sciences and Engineering, University of Southampton, Highfield, Southampton SO17, 1BJ, United Kingdom.

* corresponding author: O.Muskens@soton.ac.uk

**Abstract**

We demonstrate a new concept for reconfigurable nanophotonic devices exploiting ultrafast nonlinear control of shaped wavefronts in a multimode nanomaterial consisting of semiconductor nanowires. Femtosecond pulsed laser excitation of the nanowire mat is shown to provide an efficient nonlinear mechanism to control both destructive and constructive interference in a shaped wavefront. Modulations of up to 63% are induced by optical pumping, due to a combination of multimode dephasing and induced transient absorption. We show that part of the nonlinear phase dynamics can be inverted to provide a dynamical revival of the wavefront into an optimized spot with up to 18% increase of the peak to background ratio caused by pulsed laser excitation. The concepts of multimode nonlinear switching demonstrated here are generally extendable to other photonic and plasmonic systems and enable new avenues for ultrafast and reconfigurable nanophotonic devices.

**Introduction**

Many of the available photonic technologies are based on perfectly regular, ordered structures such as waveguides, photonic crystals, and metamaterials. There is however an increasing interest to exploit the additional degrees of freedom offered by aperiodic or disordered designs.[1-4] One way of controlling the flow of coherent energy transfer in such a medium with high efficiency is through optimization of the specific arrangement of the scatterers.[5,6] Exciting new techniques have emerged based on shaping of the light field itself to match a given scattering configuration, either through time reversal[7,8] or iterative schemes.[9,10] The method of wavefront shaping is based on the general concept that the transmission through any medium can be described by a matrix which connects all ingoing and outgoing degrees of freedom. In principle, knowledge of the transmission matrix[11,12] along with an ability to completely control the incident light[10] would allow the selection of any desired output, turning an opaque medium into a versatile optical element. Next to the interest for biomedical imaging,[13-15] wavefront shaping shows promise for reconfigurable optical elements[16-20] and control of random lasers.[21] While initial work concentrated on monochromatic continuous-wave radiation, focusing through opaque scattering media has also been achieved using ultrashort pulses[22,23] and polychromatic light.[24]

Here, we demonstrate both destructive and constructive switching of a shaped wavefront on ultrafast time scales through nonlinear optical excitation of the scattering medium. With the rapid development of applications exploiting wavefront shaping, active control of such shaped fields is of great interest. The principle is illustrated in Figure 1a. Wavefront shaping amounts to aligning the phasors resulting from independent light paths in the medium to produce a predefined output pattern, such as a single sharp focus.[10] The optimized state relies on coherence between modes exploring completely different trajectories in space and time, which can be easily perturbed by small changes to the medium. Femtosecond optical excitation of a semiconductor produces a series of nonlinear phase shifts (denoted by $\Delta\phi$ in Figure 1a) which give rise to uncorrelated, but reproducible changes in the transmission mode spectrum, as was demonstrated in our previous work.[25] In those earlier studies, the output of the medium was a random speckle pattern, limiting its use for applications. In the current work, we extend the idea of multimode dephasing by combining it with wavefront shaping to control both the destructive and constructive interference in a shaped light field on ultrafast time scale. By harnessing the output through constructive wavefront shaping, the nonlinear effects become useful and can be applied in realistic devices. Moreover, we proceed beyond dephasing by demonstrating that the dynamics can be reversed to provide a revival of a shaped wavefront on ultrafast time scales. The fact that pump-induced, nonlinear phase dynamics can be used to increase the constructive interference of a coherent superposition state was not anticipated from Ref 25 and opens up new prospects for coherent control and wavefront shaping of light in nonlinear media.

**Materials and Methods**

For our investigations we make use of mats of GaP nanowires. While previously we have used such samples for investigations of fundamental mesoscopic physics of light,[26] it is the nonlinear response of such nanowires that is of interest to this study.[25] The optical thickness of these 4.5 μm thick mats is more than 20 optical mean free paths, which is sufficient to suppress any coherent beam transmission and to produce a complete mixing of input and output degrees of freedom.[26,27]

The setup, shown in Figure 1b, combines a wavefront shaping arrangement with two-color, femtosecond pump-probe spectroscopy. A regenerative amplifier (Coherent RegA, 250kHz, 150±10 fs pulse duration) pumped optical parametric amplifier (Coherent OPA 9400) was used to provide probe and pump light with center wavelengths of 633 nm and 400 nm respectively. A 1 nm narrowband spectral filter was used to reduce the bandwidth of the probe pulses to 0.75 THz, close to the characteristic frequency correlation width of the speckle in the medium.[25,26] This allowed us to perform wavefront shaping in the spatial domain using only a single frequency mode (as opposed to temporal shaping[22,23]). As a consequence of the spectral filtering, the pulse duration of the probe was increased to 5 ps.

The pump and probe were focused onto the nanowires using a microscope objective (Nikon CFI60 LU Plan BD) with numerical aperture (NA) of 0.9. By reducing the pump collimation before the objective (L6 in Figure 1b), we achieved a homogeneous pump illumination over the entire 20-μm diameter area used for wavefront shaping. The total average incident power of the pump on the sample was 8 mW, while the used probe power was less than 0.1 mW. The pump fluence on the sample was estimated to be 15 mJ/cm$^2$, approximately 70% of the threshold at which slow, photo-induced sample degradation was observed. The time delay between probe and pump pulses was controlled using a mechanical stage, and was calibrated using the nonlinear response of a bare GaP slab without any scatterers.[25,28] For each delay position an image was taken with and without pump using a computer controlled shutter to block the pump beam in order to obtain the relative effect of the pump. Transmitted light was collected by a second, identical objective and imaged onto a monochrome 16 bit camera (AVT Stingray).

To produce an optimized spot, a digital micromirror device (DMD, Vialux) was used to control the incident probe light. We used a binary amplitude modulation scheme by Akbulut et al.[29] to switch off transmission modes that interfere destructively at a selected point, leaving only modes that interfere constructively to create a single bright focus. In the experiment, the DMD was divided into segments of 20 pixels x 20 pixels, providing a total of 900 independently controlled incident modes. The time to optimize a single segment was 30 ms, resulting in a total optimization time of 30 s. A further optimization cycle was then performed to take advantage to the improved signal to noise ratio in the optimized spot, after which no significant improvement was obtained for further cycles with a stable configuration. By positioning the focal plane of the collection objective approximately 5 μm behind the exit surface, optimizations were performed on a single spot in the far field of the scattering slab.

Images were processed using Matlab. For each data point, areas of interest (AOIs) were selected from the images corresponding to either the peak or background region. Integrated intensities for images with pump were normalized to values without pump to obtain normalized data for each time delay.

**Results and Discussion**

**Dephasing of a shaped wavefront**

Figure 2a shows the optimized spot obtained using the picosecond probe pulse, with a radially averaged cross section shown in Figure 2c (black line). An order of magnitude enhancement was obtained, which is limited by the amplitude modulation scheme combined with the finite laser bandwidth and intrinsic noise of the parametric amplifier. Ultimately, the maximum possible enhancement is set by the available number of independent transmission channels,[30] which is around 50 for the sample and illumination conditions under study.[26]

After optimization of the wavefront, the pump illumination was switched on in order to produce a nonlinear modulation of the shaped light field. Figure 2b shows the intensity map for the maximum of the pump-probe dephasing effect at 4 ps delay time, with the cross section shown in Figure 2c (red line). Figure 2d shows the pump-probe time dynamics of the transmitted intensity $T_{pump}$ normalized to its value without pump $T_{nopump}$, for the optimized spot (black line) and the average background intensity (red line). The peak to background ratio (blue line) is defined as the ratio between optimized spot and background. The main effect is a reduction of the intensity in the optimized spot down to 37% of the value without pump, corresponding to a modulation contrast of 4.3 dB. This reduction of the shaped spot is much stronger than that of the background intensity, which is reduced by only half this amount. In addition to the ultrafast effect, both the peak and the background intensities show a $T_{pump}/T_{nopump}$ ratio of around 95% for delay times less than -5 ps, due to a pump-induced heat pileup in the nanowires.

The nonlinear switching effect was found to be reproducible for different positions on the sample corresponding to different multiple scattering configurations. Figure 3a shows the normalized peak modulation for three different sample positions, while statistical results for 24 different positions are given in Figure 3b-3d. The peak to background enhancement using the binary optimization scheme varied between 8 and 16 for these optimizations (Figure 3b). The mean value of the modulation of the peak achieved at different sample positions was (50±2)% (Figure 3c), while the modulation of the background was found to be (36±2)% (Figure 3d). In addition to variations in magnitude of the switching effect, the detailed dynamics of the optimized spot were found to also depend sensitively on the local scattering configuration, as is shown in Figure 4a for three representative positions corresponding to those of Figure 3a. The variation in dynamics of the peak (labeled A-C in Figure 4a) reflects the limited number of independent transmission modes in the optimized spot.[30] As each mode is associated with a specific set of light paths through the medium, the time evolution represents the characteristic dwell times of the subset of modes that are contributing to the shaped light field for each particular configuration. Compared to the optimized spot, the integrated background intensity consists of a much larger number of transmission modes, resulting in a much smoother behavior which did not depend markedly on sample position (red lines in Figure 4a).

In absence of mesoscopic effects, the modulation of the background intensity can be described by pump-induced transient absorption during the diffuse transport time of light through the nanowire mat.[25,31] If there was only absorption, the optimized peak should follow the background and no modulation of the peak to background ratio would be observed. From the peak to background ratio, we conclude that the effect of absorption accounts for less than half of the total switching of the peak, and the remainder reflects the effect of pure dephasing processes in the switching. Further evidence of the presence of ultrafast dephasing effects is obtained by looking at the phase coherence of the background speckle. Dephasing effects result in a redistribution of intensity over the different transmission modes,[25] which is not visible when looking only at the integrated transmission intensity. Therefore, we calculated the cross-correlation *C* of the spatial intensity maps taken with ('p') and without pump ('np') at each delay time, given by

$$C_{p,np} = \frac{\sum_{x,y}[I_p(x,y) - \langle I_p \rangle][I_{np}(x,y) - \langle I_{np} \rangle]}{\sqrt{\sum_{x,y}[I_p(x,y) - \langle I_p \rangle]^2} \sqrt{\sum_{x,y}[I_{np}(x,y) - \langle I_{np} \rangle]^2}}$$

where brackets denote the average value over the selected area of interest of the image. We find a decorrelation of the speckle pattern in presence of pump illumination as shown in Figure 4b for the three positions. The time dependence of this decorrelation matches well the dynamics of the corresponding peaks A-C for each configuration, evidencing the strong role of dephasing in the modulation of the optimized wavefront.

**Ultrafast revival of a shaped wavefront**

The above results show that it is possible to selectively destroy the constructive interference of a shaped light field using the nonlinearity of the scattering medium. Our finding that nearly half of the switching is caused by pure dephasing leads to the exciting question whether the phase dynamics can be inverted to achieve constructive interference in presence of the pump. We tested the potential of such a dynamical rephasing effect in an experiment where the probe field was shaped in the presence of pump illumination and for a given pump-probe delay. Figure 5a and 5b summarize the results obtained for a range of optimizations at different pump-probe delay times. Here, the vertical black arrows represent the pump-probe delay time at which the optimization was done (the horizontal dashed lines are the baselines of the vertically shifted curves). The first thing to notice is that the background intensity (Figure 5a) is not affected at all by the optimization. This is to be expected since the background modulation only depends on transient absorption.

In comparison, the peak to background ratio (Figure 5b) shows pronounced differences between optimization with and without pump. For optimization without pump, we see for this particular configuration the suppression of the shaped field beginning around zero time delay with a maximum effect of (20±2)% at 3.5 ps.

For optimization with pump on, we find that the dynamics of the peak to background ratio strongly depend on the exact timing of the pump and probe pulses used during the optimization process. For a range of optimization delay times between -1.5 ps and 5.5 ps, the shaped field shows a clear enhancement with pump compared to without pump around the time where the optimization was done. The largest rephasing effect of (18±2)% is found for the optimization with a probe delay of 2.2 ps, corresponding to the delay at which the largest dephasing effect occurs for the optimization with pump off. Figures 5d and 5e show a more detailed analysis of the dynamics taken at the optimization at 2.2 ps probe delay. The increase of the pump to background ratio is most clearly observed in Figure 5e. This increase cannot be explained by an absorption difference between peak and background, as this would not show a maximum which shifts with the optimization time delay. We therefore attribute the relative increase of the shaped field to a coherent rephasing, driven by the pump-induced changes of the scattering medium.

We reproduced the dynamic rephasing effect on several areas of the sample as is shown in Figures 6a-6d. Figure 6a shows the intensity cross sections at the maximum rephasing time delay, normalized to the background intensity at 4 μm distance from the peak, for three independent sample positions. The full time dynamics of the rephasing effect for these different configurations are shown in Figures 6b-6d. The arrows indicate the time delay at which the optimization was done in presence of pump illumination. While nominally the same amount of maximum rephasing is observed in the peak to background ratio (blue lines), there are again variations in the dynamics depending sensitively on the sample position. As the nonlinearity will impact differently on each transmission mode, the specific dynamics in this region is likely to be an individual fingerprint of the scattering configuration. This behavior is in agreement with that observed for the dephasing effect, and shows that both the linear transmission and the nonlinear phase dynamics can potentially be engineered by a rational design of the scattering configuration.

Our results show that a large fraction of the nonlinear dephasing can be inverted to produce a coherent reshaping of the wavefront. Some losses in the inversion could be caused by nondeterministic or inelastic dephasing processes. We expect that in particular the effects of nonadiabatic modulation processes occuring on a time scale faster than the dwell time,[32,33] or reciprocity breaking[28] will produce phase dynamics and nonlinear spectral broadening which cannot effectively be harnessed for rephasing of the entire probe pulse.

**Comparison with other switching concepts**

Optical switching through multimode dephasing in a complex medium is a conceptually novel approach to all-optical control. Other semiconductor devices showing ultrafast phase modulation include photonic crystals, ring resonators and microcavities.[33-40] In these devices, typically a single resonant mode is modulated through ultrafast changes in the refractive index, resulting in a deterministic frequency shift of this mode. On the other hand, a Mach-Zehnder interferometer (MZI) is based on the phase coherence between two spatially separate pathways. Our approach can be interpreted as a multipath extension of the MZI in a highly compact configuration where many light paths are folded into a single slab of material. The spatiotemporal multimode complexity of the different paths showing different phase dynamics results in a complete reshaping of the mode spectrum. Thus, rather than a small frequency shift we see a complete collapse - or revival - of the output mode. Our device concept thus provides a switching characteristic which is very different from any other nonlinear switching device.

While differences in device geometries render a direct comparison between one, two, and three dimensional structures difficult, a simple but instructive approximation is to look at the characteristic nonlinear interaction times. The accumulated nonlinear phase $\Delta\phi$ is proportional to the time τ the light spends in the excited region according to $\Delta\phi = \Delta n \omega \tau$, with Δn the nonlinear refractive index change and ω the angular frequency. For a MZI of arm length L, the interaction time is $\tau_{MZI} = L/c$, whereas for a cavity, this time is increased to $\tau_C = FL/\pi c$, with F the cavity finesse, L the cavity length and c the speed of light. For a complex scattering medium, the average time light takes to diffuse through the excited slab of thickness L is the Thouless time $\tau_D = 3L^2/cl$, where l is the transport mean free path of light in the scattering medium.[41] For equal length L of the excited region, the scattering medium thus provides the same accumulated phase as a cavity with finesse F= $3\pi L/l$, with amounts to around 200 for our studies. Indeed, earlier experiments[25] showed a frequency shift of the overall speckle pattern of our random medium corresponding to the average accumulated nonlinear phase $\overline{\Delta\phi}$ as illustrated in Figure 1a. In comparison, a much longer device length (60 times in our example) is required for the MZI to obtain the same accumulated nonlinear phase.

In addition to the small average frequency shift, a much larger effect in our samples is caused by pure dephasing of modes.[25] Dephasing results in the decorrelation of intensity induced by a random distribution of accumulated phases proportional to $C_{p,np} \approx \langle e^{i\Delta\phi} \rangle \equiv e^{-\frac{1}{2}\langle \Delta\phi^2 \rangle}$ where the equivalence assumes a Gaussian distribution and the brackets denote an average over the ensemble of modes contributing to the intensity.[41] The decoherence model predicts a 1/e reduction of the intensity for a Gaussian distribution of accumulated phases of width $\text{var}(\Delta\phi) = 2$. The dephasing mechanism thus benefits from a broad distribution of path lengths, which is provided by the power-law distribution of diffuse transport times through the slab,[41] as well as from spatiotemporal inhomogeneity of the excitation zone. The latter condition is fulfilled by the fast time dynamics, limited penetration depth and intrinsic intensity fluctuations of pump light inside the sample, but could potentially be further optimized by structuring of the pump illumination to imprint specific nonlinear potential landscapes.[42]

**Future prospects**

While the efforts in our current work mainly focused on the proof of concept of ultrafast switching, much higher efficiencies and speeds of wavefront shaping can be easily implemented.[9-15] The use of full phase control in wavefront shaping will result in more significant enhancement in the peak to background ratios achieved. Moreover, knowledge of the transmission matrix of the material will allow generating any predetermined configuration on command within the switching time of the DMD, i.e. at kHz speeds.[11] Such speeds are of interest for a variety of applications in displays and imaging. Ultimately, we envisage that multimode devices could form the nodes of an adaptive network, where the multimode complexity provides the reconfigurability and picosecond control performs the function of all-optical communication gate. Particularly exciting is the prospect of implementing these general concepts in two dimensional plasmonic and nanophotonic waveguides, which could result in a new paradigm for reconfigurable networks.

Another application range which could take profit from the picosecond switching speed is that of ultrafast shutters for time-resolved spectroscopy. High throughput devices may be obtained by designing configurations that match the open transmission eigenchannels of the medium with order unity transmission.[10,12] A rational design of arrangements with predefined characteristics[6] will allow engineering of both the linear energy transfer and nonlinear phase dynamics similar to complex molecular systems.[43]

**Conclusion**

In conclusion we have demonstrated ultrafast optical modulation of a shaped light field by up to 63% (4.3 dB), with both dephasing and absorption performing important roles. The effect of dephasing can be partially inverted to produce a constructive revival of the light field in the presence of a femtosecond pulsed excitation. The maximum positive modulation found using this striking rephasing effect is 18%. The presence of a nonlinear scattering medium therefore adds new ways to control the transmitted light, on top of the already impressive possibilities offered by wavefront shaping.

**Acknowledgements**

The authors thank E. Bakkers from Eindhoven University of Technology for providing the nanowire sample. O.L.M. and R.B. acknowledge financial support from EPSRC through grant EP/J016918/1.

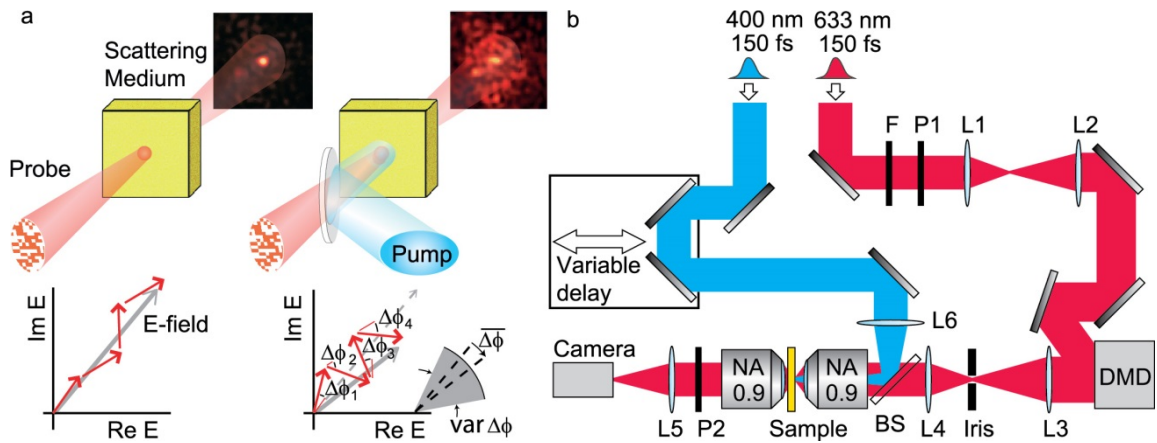

Figure 1: **Concept and experimental setup.** (a) Concept of ultrafast dephasing of shaped field. Left: wavefront shaping of the probe light using binary amplitude results in alignment of partial E-fields in complex phase plane (lines, red). Right: illumination with pump results in dephasing of shaped fields and destruction of the optimized spot. (b) Diagram of experimental setup. Lenses L1 and L2 magnify the probe by x3.3 onto the DMD, L3 and L4 demagnify the DMD pattern onto the microscope entrance by x2.5 in a 4-f configuration. L5 (f=500 mm) images the transmitted light onto the camera and L6 (f=250 mm), positioned 20 cm from the aperture of the objective, increase size of pump focus. P1 and P2 are polarizers, F is a narrowband laser line filter at 632.8 nm with a spectral linewidth of 1 nm, resulting in broadening of the probe pulse duration to 5 ps.

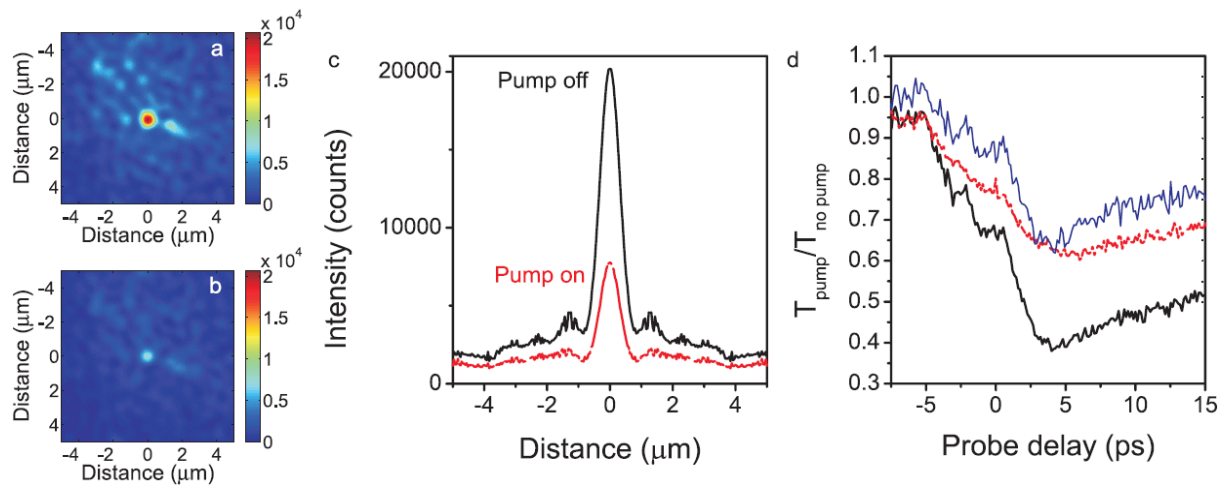

Figure 2: **Nonlinear control of optimized spot.** Images of a shaped wavefront optimized in the absence of pump illumination, for conditions 'pump off' (a) and 'pump on' (b). (c) Radially averaged cross-sections of (a,b) for pump off (line, black) and pump on (dashed line, red). The change in peak intensity is 63% with a pump delay of 2 ps. (d) Plots of the transmission in presence of the pump $T_{pump}$, normalized to value without pump $T_{nopump}$, as a function of pump delay for the peak intensity of the optimized spot (thick line, black), and the average background intensity (dashed line, red). Thin line, blue denotes peak to background ratio.

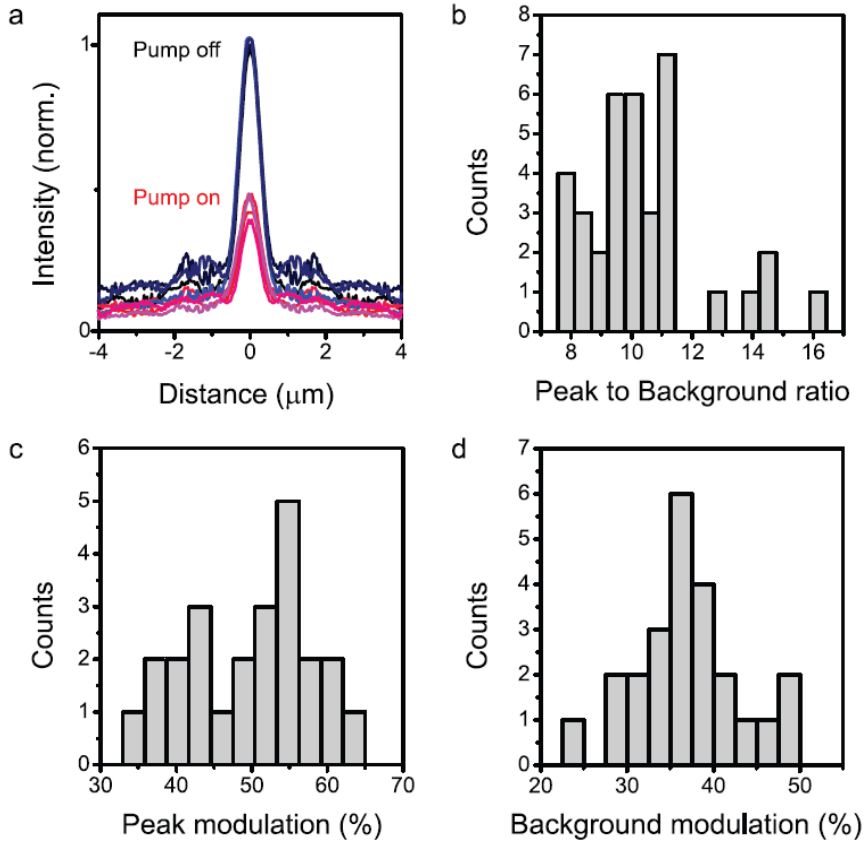

Figure 3: **Reproducibility of nonlinear switching effect.** (a) Radially averaged profiles of optimized spot without pump on three different sample positions (black-blue lines, 'pump off') and in presence of the pump (red-magenta lines, 'pump on'). Histograms demonstrating the variation in experimental results for switching a spot optimized without pump, for maximum peak optimization (b), modulation of the peak intensity (c) and integrated background away from peak (d) following excitation with pump.

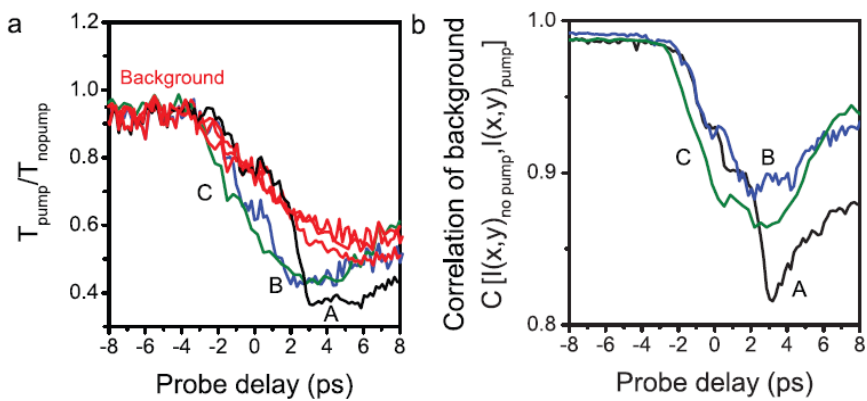

Figure 4: **Time dynamics of shaped peak and background correlation.** (a) Time dependence of the intensity of the optimized spot (black lines, labelled A-C) and average background (red lines) induced by the pump as a function of pump-probe time delay, for sample positions corresponding to the data in Figure 3a. (b) Time dependence of the cross correlation (see Methods) of the transmitted background speckle for subsequent images taken with and without pump, corresponding to sample positions A-C.

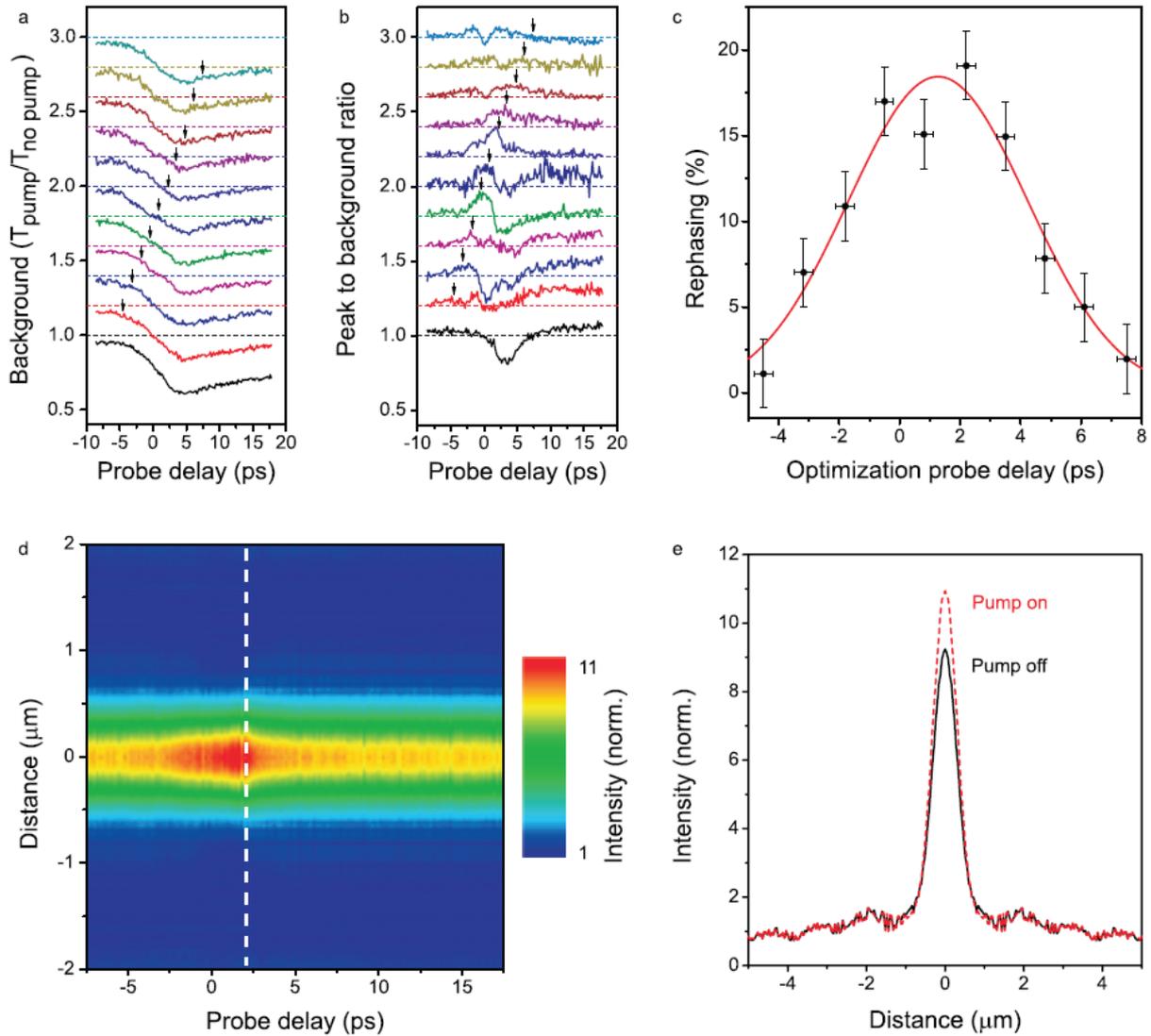

Figure 5: **Constructive dynamical reshaping of light fields.** Plots of background (BG) intensity (a) and peak to background ratio (b) for pump on, normalized to values without pump, for optimizations with pump on and varying pump-probe delays from -5.5 ps to 7.5 ps (indicated by vertical arrows). Data for optimization without pump at same spot is included for reference (black lines, bottom). (c) Plot of maximum rephasing effect at optimization point obtained from (b) as a function of probe delay (points) along with a Gaussian fit (line, red). A maximum increase in the peak to background ratio of 18% is found for optimizations carried out with a pump delay of 2.2 ps. (d) Evolution of optimized spot as a function of probe delay for an optimization carried out at 2.2 ps (indicated by dashed line). (e) Cross-section of spot optimized at 2.2 ps with (dashed line, red) and without (line, black) pump, showing the maximum positive modulation (at a delay of 1.7 ps) of the spot with respect to the background by the pump. Both (d) and (e) were radially averaged and normalized to the average background.

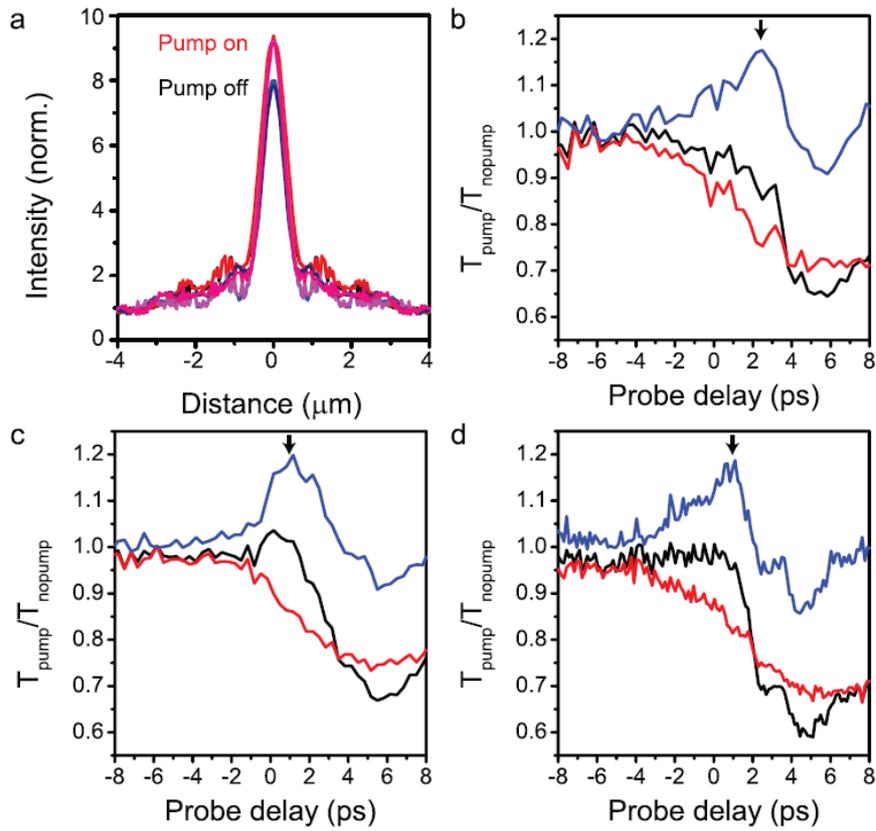

Figure 6: **Reproducibility of constructive dynamical reshaping.** (a) Radially averaged cross sections of peak optimized in presence of pump for conditions with pump (red magenta lines) and without pump (black-blue lines). (b-d) Time-dependence of the intensity of the optimized spot (black lines) and average background (red lines) for three different sample positions and for optimization with pump on at time delay given by arrows. Blue lines represent peak to background ratio, showing a dynamic reshaping of around 20% for all positions.